\documentclass[reprint, amsmath, amssymb, aps, superscriptaddress, prl]{revtex4-1}
\usepackage{graphicx}
\usepackage{dcolumn}
\usepackage{bm}
\usepackage{hyperref}

\begin{document}
  
  \title{Localization in a $t$-$J$ type ladder with translational symmetry}
  \author{Rong-Yang Sun}
  \affiliation{Institute for Advanced Study, Tsinghua University,
  Beijing, 100084, China}
  \author{Zheng Zhu}
  \affiliation{Department of Physics, Massachusetts Institute of Technology,
  Cambridge, Massachusetts 02139, USA}
  \author{Zheng-Yu Weng}
  \affiliation{Institute for Advanced Study, Tsinghua University,
  Beijing, 100084, China}
  \affiliation{Collaborative Innovation Center of Quantum Matter, Tsinghua
  University, Beijing 100084, China}
  \date{\today}

\begin{abstract}

An explicit \emph{spatial} localization of a hole is shown in a two-leg $t$-$J$
ladder in the presence of a staggered chemical potential, which still retains a
translational symmetry, by density matrix renormalization group method.
Delocalization can be recovered in the following cases, where either the hidden
phase string effect is turned off or a finite next-nearest-neighbor hopping $t'$
is added to sufficiently weaken the phase string effect. In addition, two holes
are always delocalized by forming a mobile bound pair, in contrast to the
localized single holes, which points to a novel pairing mechanism as one of the
essential properties of a doped Mott insulator.

\end{abstract}

\maketitle

\emph{\label{intro}Introduction.---}Anderson localization \cite{Anderson1958,
Abrahams1979, Lee1985} is a quantum interference phenomenon of a mobile particle
losing transport beyond a finite spatial distance as caused by random
scatterings of impurities.  Such a non-ergodic phenomenon has been recently
conjectured to persist in some interacting systems at finite energy density
known as many-body localization (MBL) \cite{Basko2006, Pal2010}. Here the
quenched disorder is an essential requirement for both cases. One may also ask
whether the localization feature can exist in an interacting system with
translational symmetry, i.e., without a quenched disorder.  The early indication
was previously discussed in the He solid \cite{YuKagan1974, Kagan1984}. \ Then a
concept called quantum disentangled liquid (QDL) \cite{Grover2014} is proposed
to describe a possible localization behavior in a system composed of two
distinct species with large mass ratio. The light species can feel an effective
random potential coming from the dynamical massive species on account of
interaction, which may then induce the localization of the light species
\cite{Schiulaz2014b, Grover2014, Papic2015, Schiulaz2015, Yao2016}.  However,
such a localization may not be stable and still show ergodicity at an
exponentially long time dubbed as quasi-MBL \cite{Papic2015, Yao2016}.  Other
attempts to realize MBL without disorder include systems with a nonlocal
interaction \cite{Mondaini2017} or with an extensive set of local conserved
quantities \cite{Smith2017b}.

On the other hand, the spatial self-localization of charges in the ground state
has recently been found \cite{He2015} by density matrix renormalization group
(DMRG) method \cite{White1993,Schollwock2005} in a Hubbard (and $t$-$J$) two-leg
ladder with very large ratio of $U/t$, where \emph{two or more holes} can be
self-trapped to form a charge droplet \cite{He2015}. Here the corresponding
superexchange coupling $J=4t^2/U$ for the spins is much smaller than the hopping
$t$ of the holes ($t/J>30$), resembling a two-species system of distinct masses.
With a relatively much reduced (still large) ratio of $U/t$ or $t/J$ ($\sim 3$),
it has been also shown that a single doped hole can exhibit another kind of
``self-localization'' \cite{Zhu2012,Zhu2015a,Zhu2015,Zhu2018} by losing its
charge to form a neutral ``spinon'', although its spatial profile may still
remain extended in finite-size DMRG calculations \cite{Zhu2012,WSK2015}. By
contrast, turning off the phase string effect \cite{Sheng1996, Wu2008} hidden in
the doped Mott insulator, both types of localization disappear in these two
cases to recover a coherent Landau quasiparticle behavior
\cite{He2015,Zhu2012,Zhu2015, Zhu2018}.  

In this paper, we report that a single hole can truly become \emph{spatially}
localized if the $t$-$J$ two-leg ladder is added by a staggered chemical
potential without breaking the translational symmetry (although the unit cell is
doubled along the ladder). Such a localized hole can be delocalized by turning
off the phase string in the so-called $\sigma\cdot t$-$J$ model with the same
staggered chemical potential. A transition to delocalization also occurs in the
$t$-$t'$-$J$ model (with the staggered chemical potential) when the
next-nearest-neighbor hopping $t'$ added is sufficiently large to diminish the
phase string effect.  Furthermore, two holes are found to form a tight-bound
pair and also become delocalized at $t/J=3$, which is in contrast to forming the
localized droplet at $t/J>30$ \cite{He2015}. It indicates that the pairing force
in such a doped Mott insulator is essentially of the same origin as the one
responsible for the localization of the single hole. 

\emph{\label{model}The models.---}The Hamiltonian is composed of three terms: $H
= H_{t} + H_{J} + H_{\mu}$, where the hopping term $H_{t} = -t \sum_{\langle ij
\rangle \sigma}c^{\dagger}_{i\sigma}c_{j\sigma} + h.c.$ and the superexchange
term $  H_{J} = J\sum_{\langle ij
\rangle}(\mathbf{S}_{i}\cdot\mathbf{S}_{j} - \dfrac{1}{4}n_{i}n_{j})$
are the same as the $t$-$J$ model, where $c_{i\sigma}$ is the annihilation
operator of the electron, $\mathbf{S}_{i}$ is the spin operator, and $n_{i}$
is the local electron density operator which is always subjected to the
no-double-occupancy condition, i.e., $n_{i} \leq 1$. The model is defined in a
two-leg ladder of square lattice  with total number of sites $N\equiv
N_{x}\times N_{y}$ (with $N_{y}=2$) \cite{Zhu2015}, and is further in the
presence of a chemical potential term $H_{\mu} = \sum_{i}\mu_{i}n_{i}$, with
$\mu_{i}=\mu_{a}$ or $\mu_{b}$ alternating for the odd and even rungs along the
one-dimensional chain direction. In the following, we shall mainly fix
$\mu_{a}=t$ and $\mu_{b}=0$ in the most calculations, but  a continuous
variation of $\mu_{a}$ will be also discussed. 
  
  \begin{figure}
    \includegraphics[width=\linewidth]{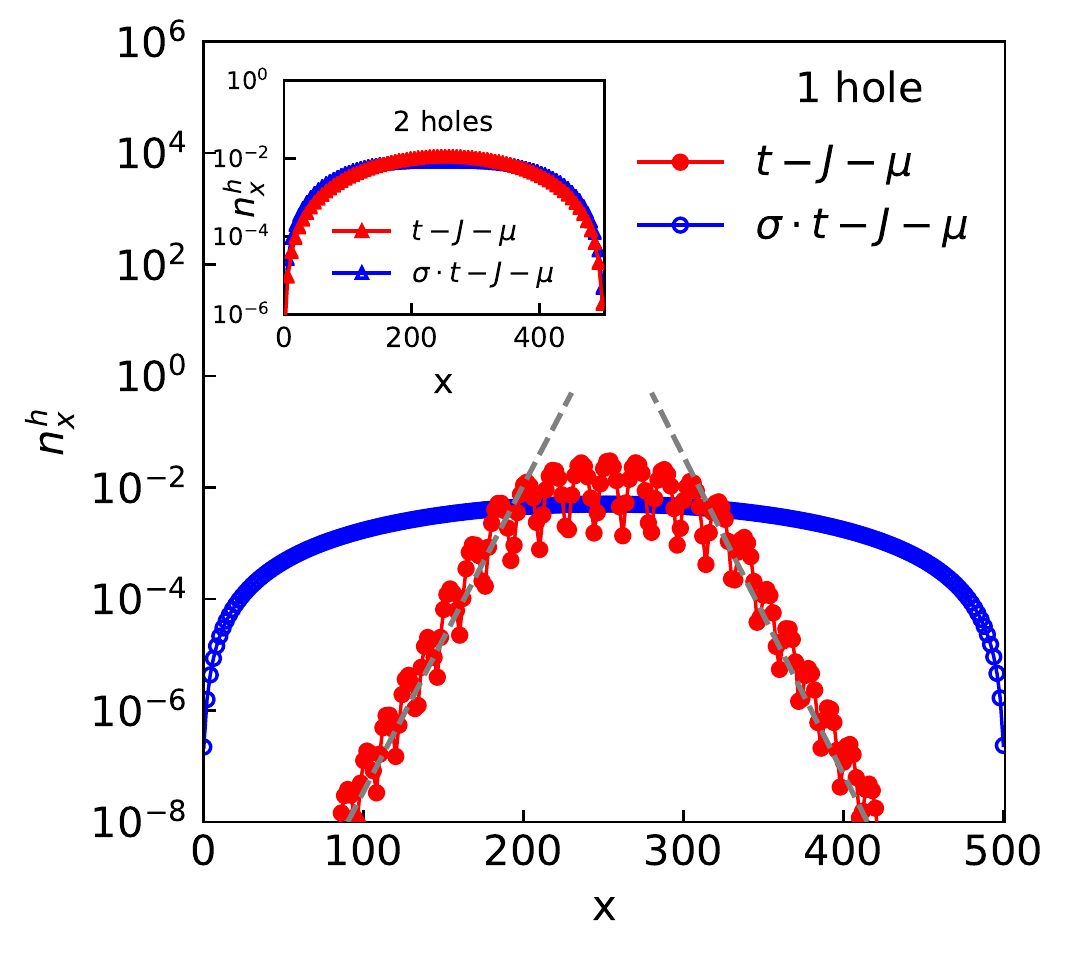}
    \caption{(Color online) The spatial distribution of the hole density defined
    on the rung labelled by $x$ in a two-leg ladder of $N=N_x\times 2$ with
    $N_x=500$ under an open boundary condition. Main panel: a hole is spatially
    localized in the central area in a $t$-$J$ model with staggered chemical
    potentials of $\mu_{a}=t$ and $\mu_{b}=0$, whereas it is always delocalized
    in the $\sigma\cdot t$-$J$ model with the same chemical potentials. Inset:
    the delocalized density profiles for two doped holes in both
    models.}\label{fig:hd}
  \end{figure}

As a comparative study, we also modify the above $t$-$J$ like model into the
so-called $\sigma\cdot t$-$J$ like model, in which only the hopping term is
changed from $H_{t}$ to $H_{\sigma\cdot t} = -t \sum_{\langle ij \rangle
\sigma}\sigma c^{\dagger}_{i\sigma}c_{j\sigma} + h.c.$ by inserting a sign
$\sigma=\pm 1$,  which can be proven \cite{Zhu2012, Zhu2015} to precisely
compensate the phase string sign structure hidden in the original $t$-$J$ model
even in the presence of $H_{\mu}$. The phase string effect is caused by the
nearest-neighbor hopping of the hole, but will get ``scrambled'' by introducing
a sufficiently large next-nearest-neighbor hopping term,  $ H_{t'} =
-t'\sum_{\langle\langle ij \rangle\rangle \sigma}
c^{\dagger}_{i\sigma}c_{j\sigma} + h.c.,$ in the $t$-$t'$-$J$ model \footnote{On
  a bipartite lattice, the phase string effect is originated from the
  nearest-neighbor hopping of a hole between the two sublattices. When a
next-nearest-neighbor hopping, $t'$, which connects the same sublattices, is
turned on, it opens up a new channel without inducing the frustrated phase
string effect. Thus adding the $t'$ term will effectively introduce the
competition to the original phase string effect that plays a dominant role in
the doped $t$-$J$ type models.}.
  
In the following DMRG  calculation, we fix $t/J =
3$ as the same in \cite{Zhu2012,Zhu2015,Zhu2013} and focus on the $\mu_{a}=t$
and $\mu_{b}=0$ to study the ground states of the one hole and two hole doped
cases. At $\mu_{a}=\mu_{b}=0$, we recover the isotropic 2-leg $t$-$J$ ladder
results \cite{Zhu2012,Zhu2015,Zhu2013}. In the simulation, we choose an open
boundary condition (OBC) along the chain direction and keep 300 to 500 states to
control the truncation error up to the order of $10^{-11}$ for one hole doped
case and $10^{-10}$ for the two hole case with 200 to 3000 sweeps.

\emph{\label{hdd}Hole density distribution.---}Define the hole density per rung
by summing up two sites of each rung labelled by $x$:
  \begin{equation}
    \label{eq:hd-rung}
    n^{h}_{x} \equiv \sum_{i \in  x} (1 - n_{i})~.
  \end{equation}
Its distribution is presented in the main panel of Fig. \ref{fig:hd} for the
single-hole ground states of the $t$-$J$ (red dots) and $\sigma\cdot t$-$J$
(blue dots) ladders in the presence of an alternating chemical potential,
$\mu_{a}=t$ and $\mu_{b}=0$, as a function of $x$  along the one-dimensional
chain direction.

The most striking feature illustrated in Fig. \ref{fig:hd} is that the single
hole is clearly localized in the central area of the ladder in the staggered
$t$-$J$ model (red dots). Here an exponential fall off by six orders of
magnitude in the hole density away from the central region merely spanned by
about $150$ rungs is shown. This is in sharp contrast to the delocalization
profile of the single hole in the $\sigma\cdot t$-$J$ model, which only differs
from the  $t$-$J$ model by a sign $\sigma$ in the hopping term, with the same
superexchange term and staggered chemical potentials. 

Furthermore, the density distributions of two doped holes are presented in the
inset of Fig. \ref{fig:hd}, where the charge density profiles become delocalized
in both models. This is in sharp contrast to the single-hole case of the
staggered $t$-$J$ model, where the hole is well localized spatially in the main
panel. It implies that two holes must form a new object to become mobile, which
is to be further examined below.

  \begin{figure}
    \includegraphics[width=\linewidth]{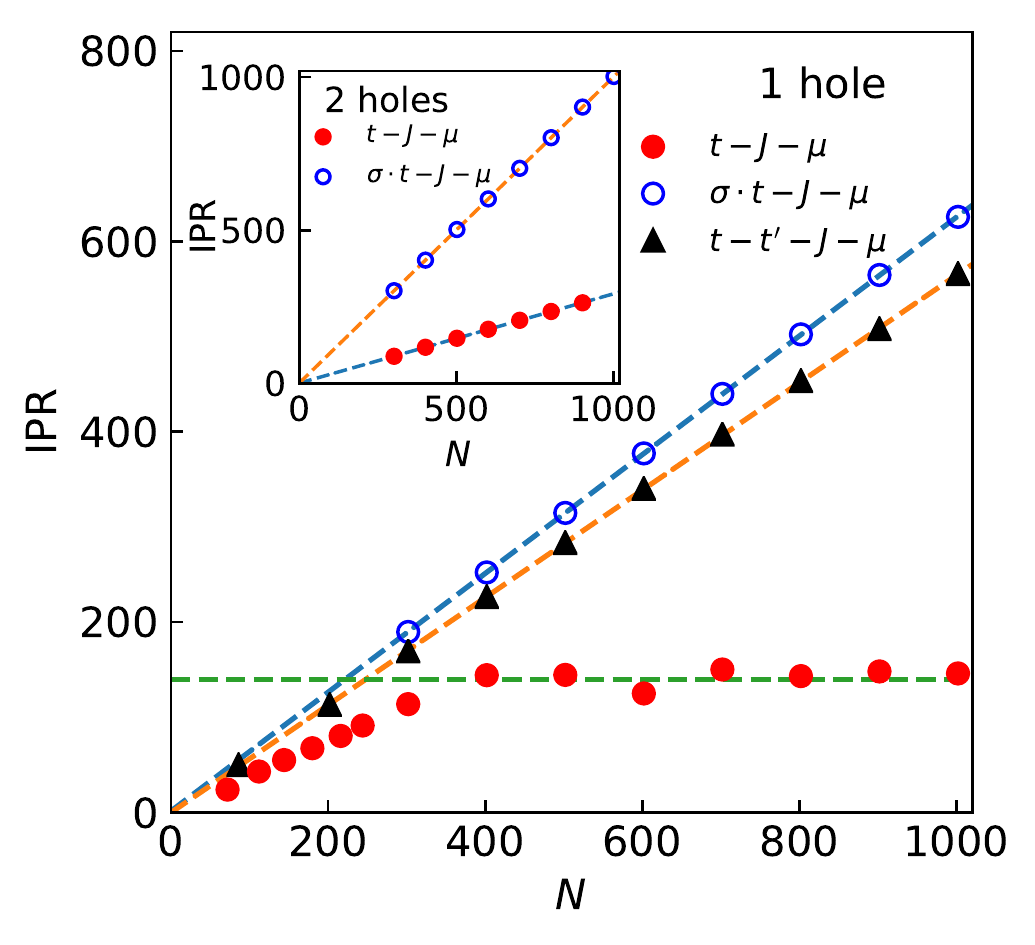}
    \caption{(Color online) The spatial localization of the hole as measured by
    IPR in the staggered $t$-$J$ model (red full cycles). By contrast, the
    linear $N$-dependences of the IPR for the stagger $\sigma\cdot t$-$J$
    model (open circles) as well as the $t$-$t'$-$J$ model at $t'/t=0.3$ (full
    triangles) indicate the delocalization of the single hole. Inset:  IPR
    $\propto N$ for the delocalized two-hole cases. }\label{fig:1hipr}
  \end{figure}

\emph{\label{1hipr}Inverse participation ratio for the one hole case.---}A
useful quantity to measure the localization in the single particle theory is the
inverse participation ratio (IPR) defined by
  \begin{equation}
    \label{eq:ipr-1h}
    \mathrm{IPR} = \dfrac{1}{\sum_{i} p_{i}^{2}}~,
  \end{equation}
where $p_{i}$ is the probability that the single particle is located at site $i$
satisfying $\sum_i p_i=1$. To generalize this concept to a many-body system with
one doped hole, we may define $p_{i}$ as the probability that the hole is at
site $i$ by tracing out the spin background. The single hole ground state can be
expressed as
  \begin{equation}
    \label{eq:1hstate}
    |\Psi_G\rangle = \sum_{i,\{\sigma\}_{i}} C_{i,\{\sigma\}_{i}} |i,\{\sigma\}_{i}\rangle~,
  \end{equation}
where $|i,\{\sigma\}_{i}\rangle$ denotes the Ising basis $\{\sigma\}_{i}$ with
one hole at site $i$. Then $p_{i} \equiv \sum_{\{\sigma\}_{i}}
|C_{i,\{\sigma\}_{i}}|^{2}=1-n_{i}$ . 
  
  \begin{figure}
    \includegraphics[width=\linewidth]{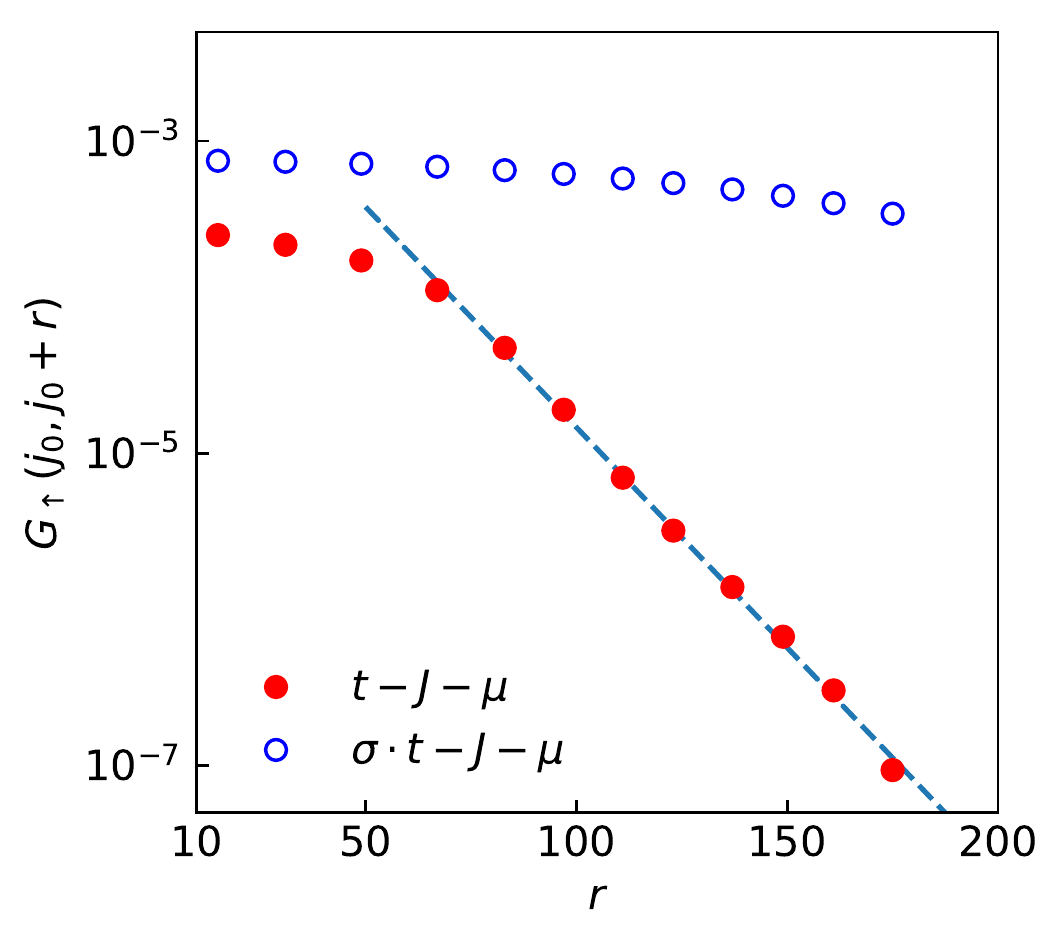}
    \caption{(Color online) The single-hole equal-time propagator in the
      one-hole ground state, which decays exponentially at large distance for
      the staggered $t$-$J$ model (red full circles) as opposed to an extended
      motion in the staggered $\sigma\cdot t$-$J$ model (blue open circles).
    Here $N_x=500$. }\label{fig:cdagc-cor}

  \end{figure}
  
If the hole is localized at one site with $p_i=1$, one finds $\mathrm{IPR}=1$
according to Eq. (\ref{eq:ipr-1h}). In the opposite limit that the hole is
uniformly distributed over $N$ sites of the lattice in a delocalized state,
$n^{h}_{i} = 1/N$ or $\mathrm{IPR}=N$, which is linearly scaling with $N$. The
IPR versus $N$ in the present staggered $t$-$J$ and staggered $\sigma\cdot t$-$J$
ladders are shown in the main panel of Fig. \ref{fig:1hipr}. For the staggered
$t$-$J$ model, initially $\mathrm{IPR}$ increases linearly with smaller $N$ but
then quickly saturates to a constant as $N=N_x\times 2$ becomes larger than
$400$ (red dots). Such a scaling behavior is consistent with the localization
picture previously found in Fig. \ref{fig:hd}. On the other hand, $\mathrm{IPR}$
remains linearly proportional to $N$ for the $\sigma\cdot t$-$J$ model (blue
open circles), also consistent with the delocalization behavior in Fig.
\ref{fig:hd}. 

Furthermore, later we shall discuss a localization-delocalization transition by
adding a next-nearest-neighbor hopping $t'$ at a critical point $t'/t=t'_c/t\sim
0.15$, and in Fig. \ref{fig:1hipr} the corresponding $\mathrm{IPR}$ clearly
exhibits a linear $N$ or delocalized behavior (black triangles) at $t'/t=0.3>
t'_c/t $.

In the inset of Fig. \ref{fig:1hipr}, the IPRs for the two hole cases of the
staggered $t$-$J$ (red full circles) and staggered $\sigma\cdot t$-$J$ (blue
open circles) models indicate linear-$N$ scaling behaviors, which support the
delocalization picture of two holes in both cases. Here the probability for
finding a hole at site $i$, $p_i\equiv  (1-n_i)/2$, is defined for a two hole
case.
  
\emph{\label{etgf}Equal-time one-hole propagator.---}The equal-time hole
propagation in the \emph{single-hole} ground state $|\Psi_{G}\rangle$ is defined
as
  \begin{equation}
    \label{eq:etgf}
    G_{\sigma}(i, j) \equiv \langle \Psi_{G}| c^{\dagger}_{i\sigma}
    c_{j\sigma}|\Psi_{G}\rangle~.
  \end{equation}
Note that this is not a conventional equal-time Green's function in which
$|\Psi_{G}\rangle$ is taken as the half-filling ground state. The quantity
defined in Eq. (\ref{eq:etgf}) tracks the motion of the hole in its true ground
state. Define the distance between site $i$ and $j$ on the same chain of the leg
by $r\equiv |i-j|$, with the reference point $j=j_0$ fixed at the central site,
the calculated quantity is shown in Fig. \ref{fig:cdagc-cor}.

$G_{\uparrow}(j_{0}, j_{0} + r)$ indicates an exponential decay at large $r$ in
the staggered $t$-$J$ model (red full circles), but it remains finite in the
staggered $\sigma\cdot t$-$J$ system (blue open circles). It is consistent with
the previous results indicating that the hole is localized in the former as
opposed to its delocalization behavior in the latter. 

\emph{Localization-delocalization transitions.---}So far we have established by
the DMRG calculation that a single hole doped into a two-leg quantum spin ladder
will be spatially localized in the $t$-$J$ model in the presence of a staggered
chemical potential, $\mu_{a}=t$ and $\mu_{b}=0$, along the chain ($x$) direction
of the ladder. Note that the latter has doubled the unit cell along the
$x$-direction but does not truly break the translational symmetry. By contrast,
by turning off the phase string effect with inserting a sign factor $\sigma=\pm
1$ into the hopping term of the staggered $t$-$J$ model, one obtains the
so-called staggered $\sigma\cdot t$-$J$ model, in which the localization of the
hole is immediately removed to result in a delocalized single hole ground state.

 \begin{figure}
    \includegraphics[width=\linewidth]{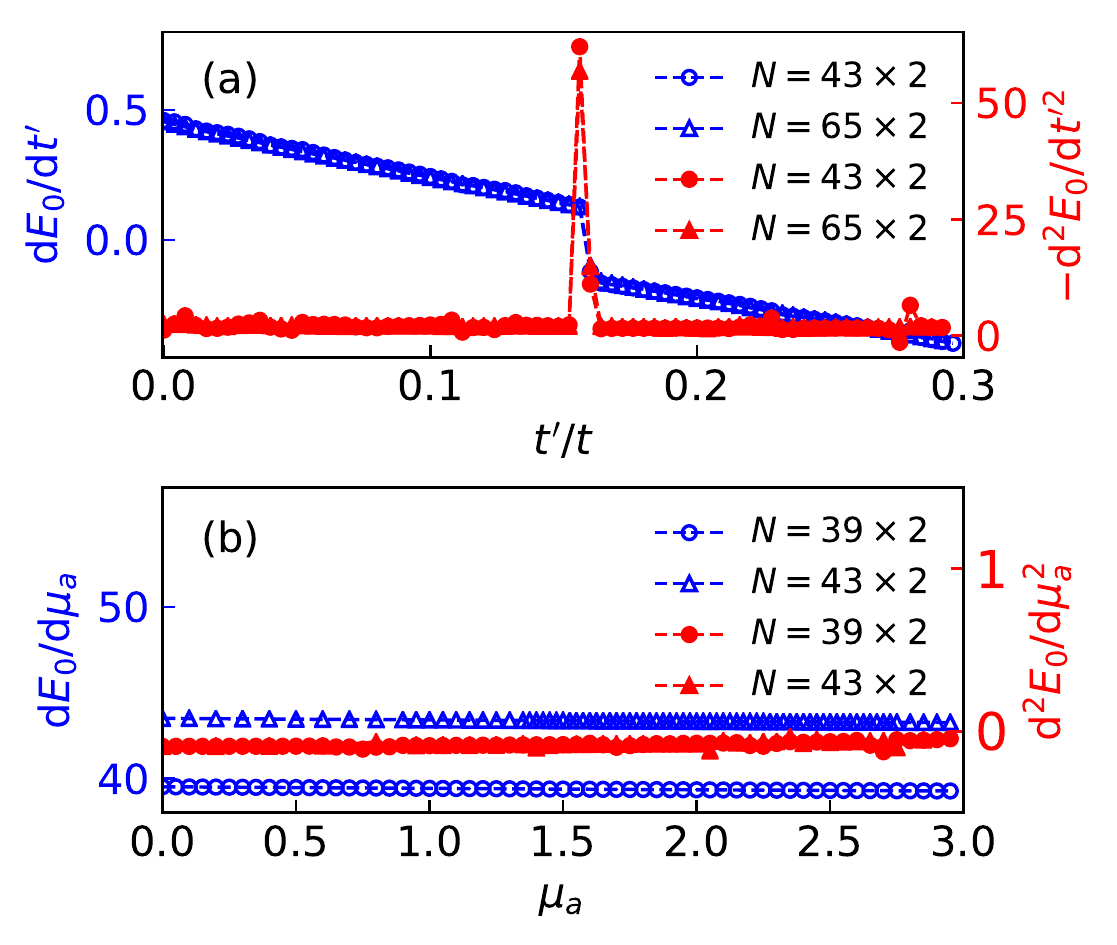}
    \caption{(Color online) (a) The first-order derivative (open symbols) and
      the second-order derivative (full symbold) of the ground state energy over
      the next-nearest-neighbor hopping $t'$ shows a ``phase transition'' at
      $t'/t \sim 0.15$ in the staggered $t$-$t'$-$J$ model; (b) The first-order
      derivative (open symbols) and the second-order derivative (full symbols)
      of the ground state energy over $\mu_{a}$ imply an adiabatic continuity in
      the one-hole state of the staggered $t$-$J$ model.
    }\label{fig:E0_vs_mu_and_tpr_div}
  \end{figure}

Instead of inserting a sign factor to switch off the phase string in the $t$-$J$
model, one may add the next-nearest-neighbor hopping $t'$ to the model, which
can also suppress the phase string effect. It is expected that when the ratio
$|t'/t|$ is large enough, the phase string gets sufficiently reduced to restore
the delocalization in such a translationally invariant system. Indeed in Fig.
\ref{fig:E0_vs_mu_and_tpr_div}(a), a ``second-order-like'' phase transition as
clearly indicated by a discontinuity in the first-order derivative and a
divergent-like sharp peak in the second-order derivative of the ground state
energy versus $t'$ at a critical point $t'_{c}/t\sim 0.15$ (we focus on $t'/t>0$
side). Correspondingly the delocalization as IPR $\propto N$ is indeed shown in
the main panel of Fig.  \ref{fig:1hipr} at $t'/t=0.3> t'_c/t $ (black
triangles).

On the other hand, if we focus on the staggered $t$-$J$ ladder and continuously
reduce $\mu_{a}$ from $\mu_{a}=t$ to the uniform limit: $\mu_{a}=\mu_{b}=0$,
there is no phase transition as shown in Fig. \ref{fig:E0_vs_mu_and_tpr_div}(b),
with the continuous and smooth behavior in the first and second order
derivatives of the ground state energy versus $\mu_{a}$. It implies that at
least in finite-size systems with $N\sim 40\times 2$, which are much larger than
the internal size \cite{Zhu2015} of the doped hole,  the single hole state in
the uniform $t$-$J$ ladder with $\mu_{a}=\mu_{b}=0$ is in the same phase as that
of the staggered $t$-$J$ ladder. Note that in the uniform $t$-$J$ ladder,
although the single hole density profile has been previously shown to be
extended in finite-size calculations\cite{Zhu2012,WSK2015}, the charge response
to a magnetic flux inserting into the ring formed by the ladder has been indeed
shown to vanish exponentially with the circumference of the ring or the ladder
length $N_x$ \cite{Zhu2012,Zhu2015}. Furthermore, the ground state is shown
\cite{Zhu2018} to have an intrinsic translational symmetry breaking as composed
of two components, with a Bloch-like standing wave superposed on top of an
incoherent component induced by the randomness of the phase string. Thus, the
strong staggered chemical potential in the present case may significantly
enhance the scattering between the two components to result in a true spatial
self-localization.

  \begin{figure}
    \includegraphics[width=\linewidth]{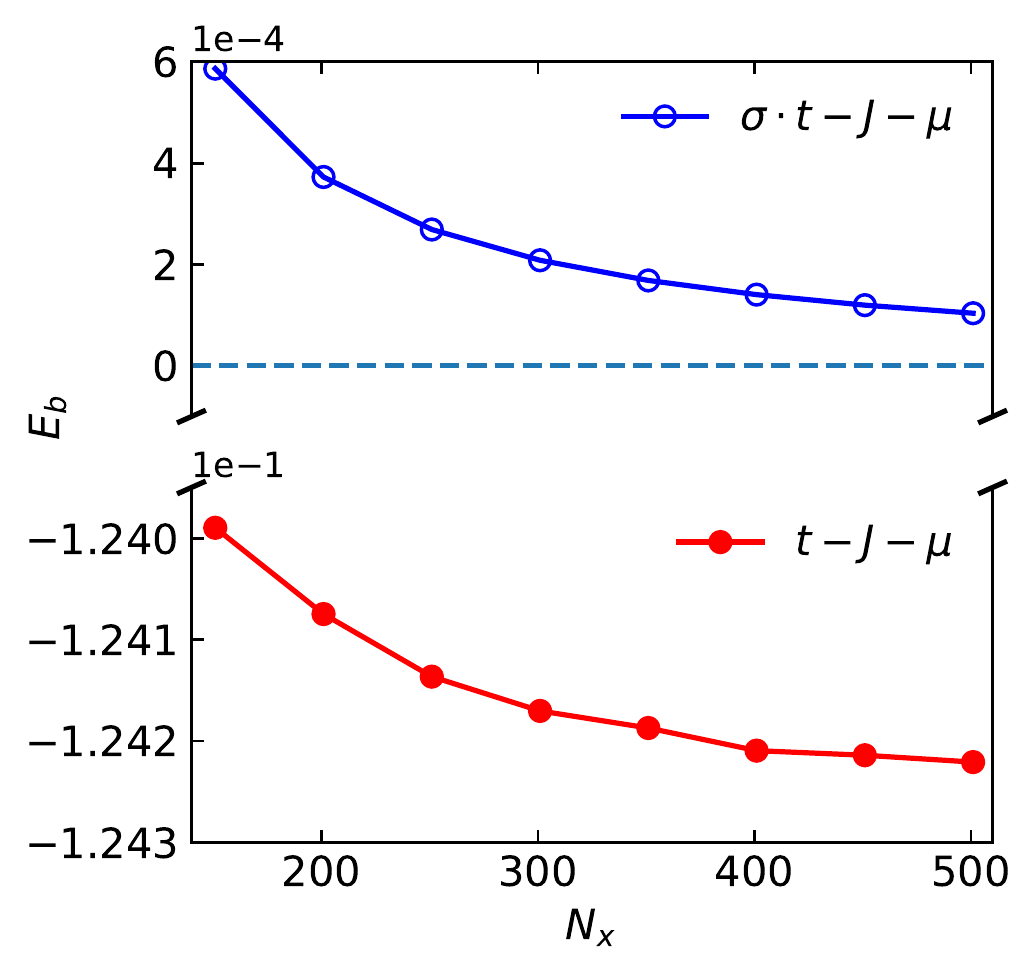}
    \caption{(Color online) The binding energy $E_b$ for two holes, which
    saturates to a negative value $\sim 0.12J$ in the staggered $t$-$J$ case,
    but approaches zero in the staggered $\sigma\cdot t$-$J$ case.}
    \label{fig:2hipr_be}
  \end{figure}

\emph{Pairing mechanism.---}In contrast to the localization of a single hole in
the staggered $t$-$J$ model, two holes have exhibited delocalization behavior in
the density profile and IPR scaling, as shown in the insets of Figs.
\ref{fig:hd} and \ref{fig:1hipr}, respectively. In the following, we show that
such two doped holes are actually paired up to become a mobile object freely
moving on a gapped spin background just like the case at $\mu_{a}=\mu_{b}=0$
\cite{Zhu2013}. 

Define the binding energy of two injected holes by \cite{Zhu2013}
  \begin{equation}
    \label{eq:be}
    E_{b} \equiv \left(E^{2h}_{0}-E^{0h}_{0}\right)-2\left(E^{1h}_{0}-
    E^{0h}_{0} \right),
  \end{equation}
where $E^{0h}_{0}$ is the ground state energy of the undoped spin ladder at
half-filling, $E^{1h}_{0}$ is the ground state energy of the one hole doped
system and $E^{2h}_{0}$ is the ground state energy of the two hole doped system.
If $E_{b} < 0$, two doped holes will form a bound pair to lower the total
energy.

The binding energies for the staggered $t$-$J$ (full red circles) and staggered
$\sigma\cdot t$-$J$ (open blue circles) models as a function of the length
$N_{x}$ of the two-leg ladder are shown in Fig. \ref{fig:2hipr_be},
respectively.  One sees a negative $E_{b}$ in the staggered $t$-$J$ case,
clearly indicating that the two holes are paired up with $E_{b}\sim -0.12 J$ at
large $N_{x}$. In contrast, the positive binding energy $E_{b}$ is shown in the
staggered $\sigma\cdot t$-$J$ model at finite $N_{x}$, which is expected to be
extrapolated to $0$ as $N_{x}\rightarrow \infty$. Namely, despite that two holes
are always delocalized in both models (cf. the insets of Figs.  \ref{fig:hd} and
\ref{fig:1hipr}), they must form a bound pair in the staggered $t$-$J$ model in
order to overcome the tendency of localization for the individual holes in this
model. It therefore points to a novel pairing mechanism of kinetic-energy-driven
\cite{Zhu2013}, which is however not present in the staggered $\sigma\cdot
t$-$J$ model even though the superexchange term $H_J$ remains the same.
 
\begin{acknowledgements}

Stimulating discussions with D. N. Sheng, H.-C. Jiang, Q.-R. Wang, and S. Chen
are acknowledged. This work is supported by Natural Science Foundation of China
(Grant No. 11534007), MOST of China (Grant No. 2015CB921000, 2017YFA0302902).

\end{acknowledgements}


\end{document}